\documentstyle[preprint,aps,prl,epsfig]{revtex}
\tightenlines

\newcounter{listfig} 

\begin{document}


\title{Ferromagnetic Ga$_{1-x}$Mn$_x$N epilayers versus antiferromagnetic GaMn$_3$N clusters}
\author{R. Giraud, S. Kuroda$^*$, S. Marcet, E. Bellet-Amalric, X. Biquard, B. Barbara$^1$, D. Fruchart$^2$, D. Ferrand, J. Cibert and H. Mariette}
\address{CEA-CNRS group `Nanophysique et Semi-conducteurs', Laboratoire de Spectrom\'etrie Physique, Universit\'e Joseph Fourier, and DRFMC-SP2M, CEA Grenoble, 38054 Grenoble, France\\
$^1$ Laboratoire de Magn\'etisme Louis N\'eel, CNRS, BP166, 38042 Grenoble, France\\
$^2$ Laboratoire de Cristallographie, CNRS, 38042 Grenoble Cedex-09, France\\
$^*$ On leave from Institute of Materials Science, Univ. of Tsukuba, Japan}

\date{\today}
\maketitle

\begin{abstract}

Mn-doped wurtzite GaN epilayers have been grown by nitrogen plasma-assisted molecular beam epitaxy. Correlated SIMS, structural and magnetic measurements 
show that the incorporation of Mn strongly depends on the conditions of the growth. Hysteresis loops which persist at high temperature do not appear to 
be correlated to the presence of Mn. Samples with up to 2$\%$ Mn are purely substitutional Ga$_{1-x}$Mn$_x$N epilayers, and exhibit paramagnetic properties. At 
higher Mn contents, precipitates are formed which are identified as GaMn$_3$N clusters by x-ray diffraction and absorption: this induces a decrease of the 
paramagnetic magnetisation. Samples co-doped with enough Mg exhibit a new feature: a ferromagnetic component is observed up to $T_c\sim175$~K, 
which cannot be related to superparamagnetism of unresolved magnetic precipitates.

\end{abstract}
\pacs{75.50.Pp, 71.55.Eq, 61.10.Ht}


\narrowtext

Since the discovery of giant magneto-resistance in ferromagnetic metal multilayers \cite{Baibich88}, spin-dependent phenomena rapidly appeared as a 
unique mean to merge information storage and processing, thus rising one of the powerful concepts of a spin-based electronics \cite{Wolf01}. 
Nevertheless, metal-semiconductor interfaces could hinder the injection of a spin-polarized current into a semiconductor \cite{Schmidt00}, although an 
additional tunneling barrier is predicted to overcome this issue \cite{Rashba00,Fert01}.  
Diluted magnetic semiconductors (DMS -see, e.g., \cite{Ohno98,Ferrand01}-) showing room-temperature ferromagnetism appear as a serious alternative to ferromagnetic metals to ensure an efficient 
injection and, indeed, high critical temperatures $T_c$ were predicted in Mn-doped wide band-gap semiconductors \cite{Dietl00}. 
As a result, there is a growing interest in the search for DMS having both a high $T_c$ and a 
large spin polarization of the carriers, which could then be used as spin polarizers or filters and for spin injection into a nonmagnetic semiconductor 
(similarly to other DMS which were used at low temperature \cite{Fiederling99,Ohno99}). Room-temperature ferromagnetism was previoulsy reported in 
GaN:Mn thick films \cite{Sonoda02,Reed01,Thaler02}, with an extrapolated $T_c \sim 940$~K \cite{Sonoda02}. 
However, these results remain ambiguous: a large additional paramagnetic component is also observed \cite{Zajac01,Soo01,Ando03}, and ferrimagnetic clusters, such as Mn$_4$N precipitates with the perovskite structure, 
could be at the origin of a ferromagnetic-like behaviour \cite{Cui02,Rao02,Kim03}. 

In order to determine the specific features of the pure Ga$_{1-x}$Mn$_x$N phase, we investigated in detail the structural 
and magnetic properties of GaN:Mn thick epilayers with the wurtzite structure. These were grown by nitrogen-plasma assisted molecular beam epitaxy (MBE) on a GaN buffer layer,  
previously elaborated by metal-organic chemical vapor deposition on a sapphire substrate. 
The manganese content was always checked by secondary ion mass spectrometry (SIMS), using an Mn-implanted GaN layer as a reference: 
depth profiles were homogeneous over the 0.3~micron layer thickness. 
As discussed below, no evidence of any secondary phase was found for Mn incorporations as large as 2$\%$, using extended x-ray absorption spectroscopy (EXAFS) 
as a local probe of the Mn environment, whereas we give direct evidences for the formation and the nature of antiferromagnetic precipitates with the perovskite structure 
above 2$\%$. Every sample exhibits paramagnetism at low temperature, as well as a smaller hysteretic artefact persisting at high temperature which is 
not specific to our epilayers. Moreover, the low-field susceptibility also reveals the coexistence of magnetic phases in diluted Ga$_{1-x}$Mn$_x$N epilayers, 
if substantially co-doped with Mg, a widely used acceptor in GaN. This behaviour is not due to clusters of another phase, and it may suggest the occurence of hole-induced 
ferromagnetism, but still with a $T_c$ lower than 300~K.

\begin{figure}     
\centerline{\epsfxsize= 18cm \epsfbox{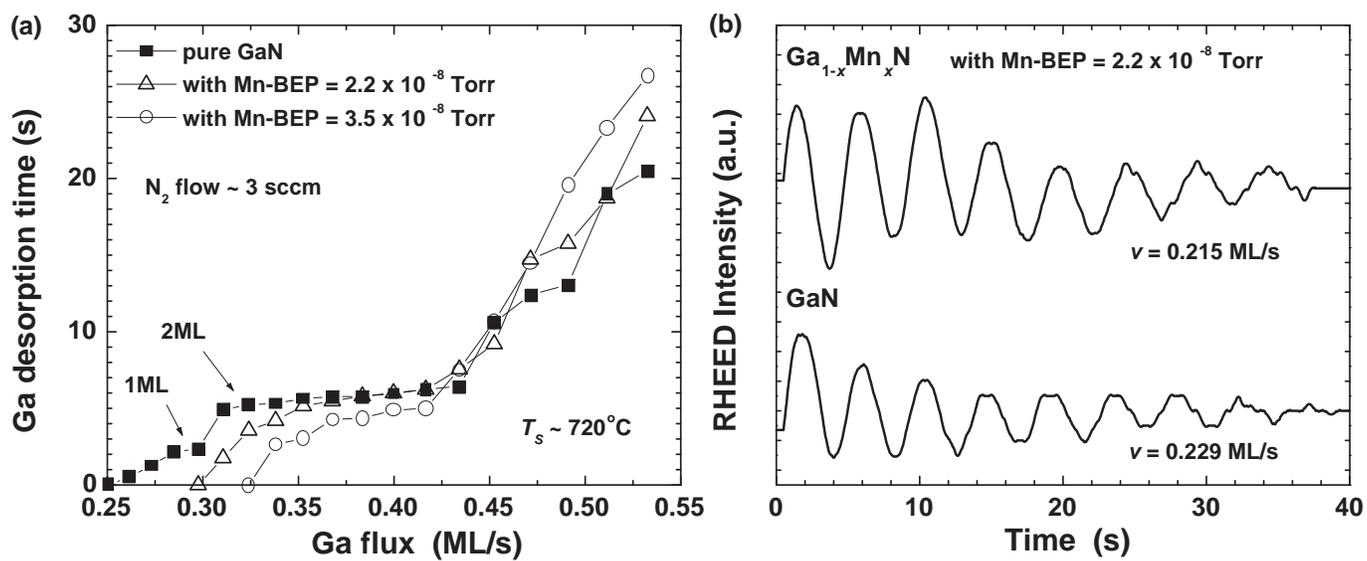}}
\caption{R. Giraud \emph{et al.}}
\label{fig1}
\end{figure}

We first focus onto the growth of Mn-doped GaN epilayers without any additional doping. 
The MBE-growth of GaN was previously studied in detail \cite{Heying00,Adelmann02}. It was shown that the growth front of GaN strongly depends on the 
Ga/N flux ratio. Under N-rich conditions, a rough N-terminated surface is formed, which is stable when growth is stopped. Under Ga-rich conditions, 
a wetting Ga film is formed, which can be evidenced by the dwell time of the reflection high-energy electron diffraction (RHEED) transient signal 
during the growth interruption: a thorough calibration shows that this dwell time is a measure of the Ga-film desorption \cite{Adelmann02}. At low Ga excess, the Ga surface 
coverage increases with the Ga flux. At intermediate Ga flux, a plateau is observed (fig.~\ref{fig1}a), which corresponds to the formation of a self-regulated 
bilayer of Ga at the growth front. This smooth surface stabilizes a 2D growth of GaN within a finite growth window so that, the growth 
is fairly insensitive to fluctuations of the Ga flux and of the substrate temperature. 
The addition of a Mn flux reduces both the width of this plateau (fig.~\ref{fig1}a), and the growth rate under N-rich conditions (fig.~\ref{fig1}b):
both effects point to a reduction of the Ga sticking coefficient, which tends to limit the self-regulated 2D growth of Mn-doped GaN epilayers. 
Finally, at even larger Ga flux, Ga droplets are formed, as evidenced by the increase of the Ga desorption time (fig.~\ref{fig1}a). 
It is important to have a precise knowledge of this diagram (fig.~\ref{fig1}a), since the Mn incorporation, for a given Mn flux, strongly depends on the growth 
regime \cite{Kuroda03}. A significant amount of manganese, up to 18$\%$, was  incorporated under N-rich conditions, close to stoichiometry, whereas the 
incorporation of Mn was reduced under Ga-rich conditions within the self-regulated growth window: for a given, small-enough, Mn flux, the Mn content is 
found to be ten times smaller under Ga-rich conditions as compared to the N-rich case. This behaviour may come from a decrease of the Mn sticking coefficient 
when the Ga surface coverage is increased or it may mean that the Ga bilayer acts as a diffusion barrier 
hindering the incorporation into GaN.  

\begin{figure}     
\centerline{\epsfxsize= 18cm \epsfbox{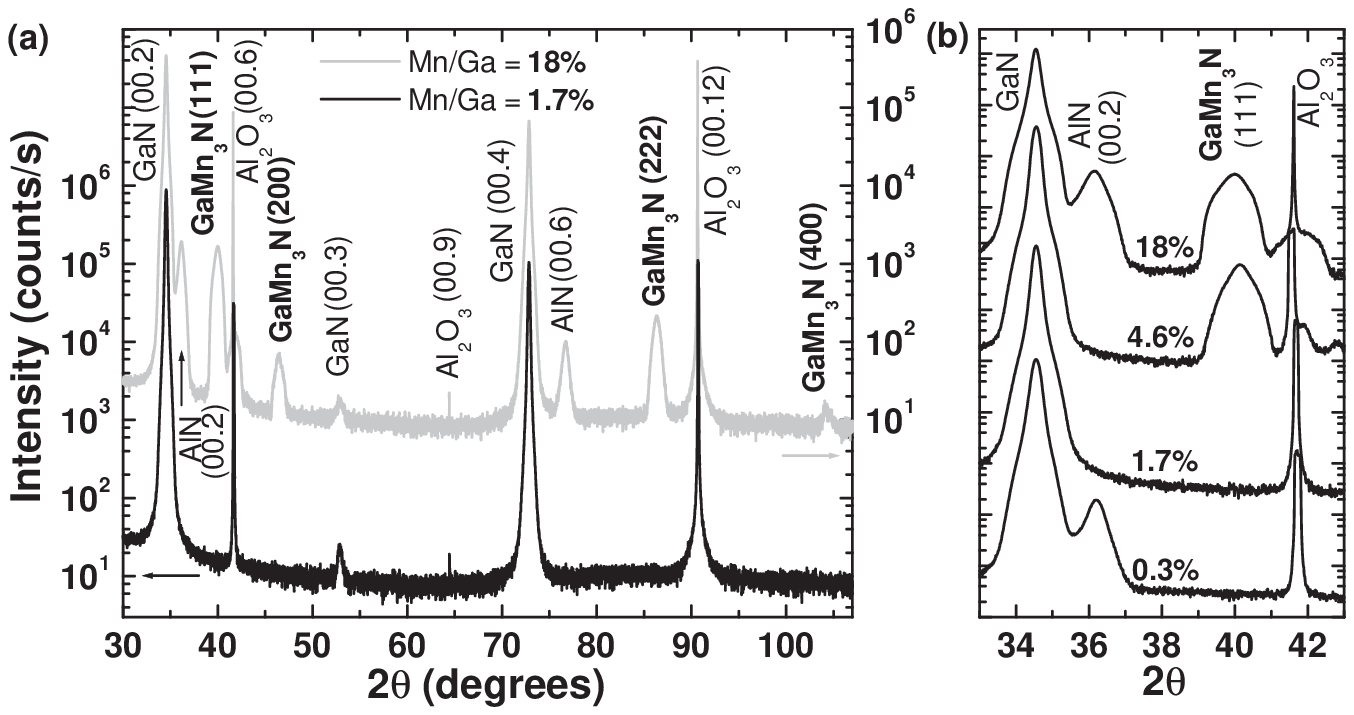}}
\caption{R. Giraud \emph{et al.}}
\label{fig2}
\end{figure}

The epilayers structure was studied using a SEIFERD 3003 PTS-HR x-ray diffractometer. The measurements do not show any spurious phase for a Mn content smaller than 2$\%$, whereas the formation of additional 
clusters with the perovskite structure is evidenced at larger doping levels (see fig.~\ref{fig2}). These could be either 
GaMn$_3$N or Mn$_4$N clusters: as these two phases have nearly the same lattice parameter \cite{Bouchaud68}, one can hardly distinguish between them by x-ray diffraction. 
These precipitates show two crystalline orientations, and most have their [111] direction (three-fold axis) colinear to the 
wurtzite $c$-axis. Note that, using a vicinal silicon surface ($\sim4^\circ$ off) as a sample holder to achieve a low background, the large dynamical range allows us to evidence the presence of these oriented clusters for a Mn-content 
down to about 2$\%$ of Mn, instead of 6$\%$ in \cite{Kim03}. Although a small amount of tiny clusters cannot be excluded (less than a few percents, because of the measurement accuracy given 
by our small but finite background), the higher sensitivity of x-ray absorption measurements also allows us to demonstrate the absence of such precipitates 
below about 2$\%$ of Mn.

\begin{figure}     
\centerline{\epsfxsize= 18cm \epsfbox{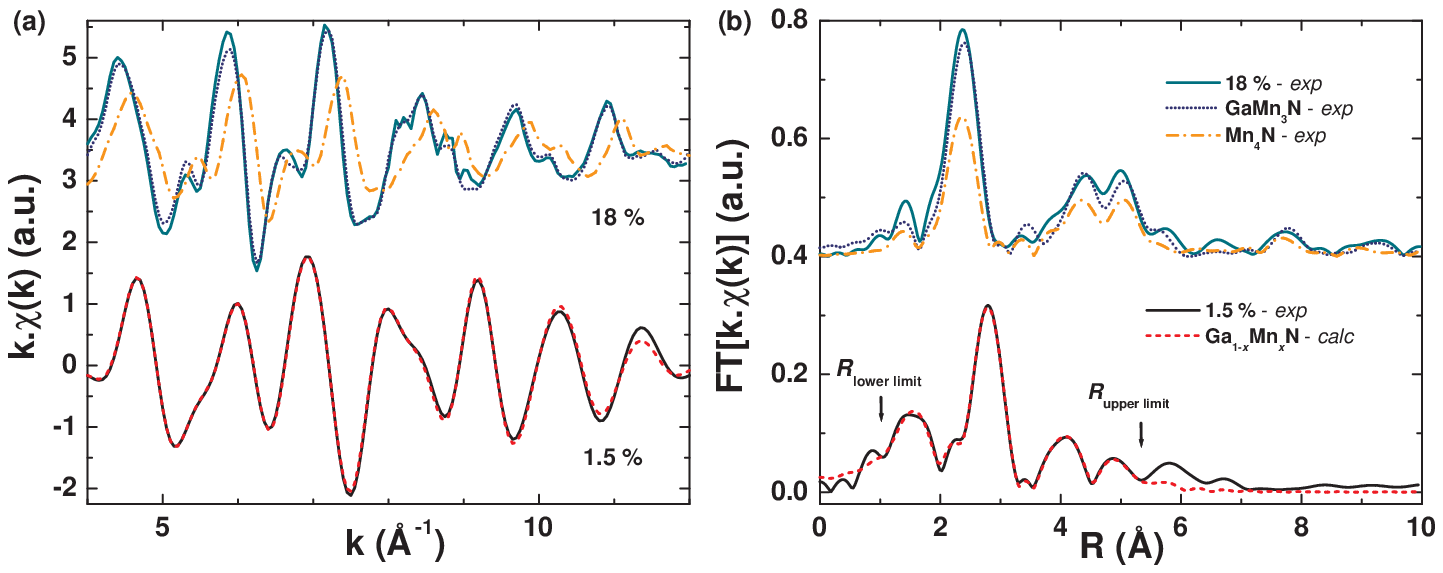}}
\caption{R. Giraud \emph{et al.}}
\label{fig3}
\end{figure}

We performed EXAFS measurements at the Mn K-edge on both our GaN:Mn epilayers and two perovskite powders, at the European Synchrotron Radiation Facility. 
This analysis was intended, first to completely discard the occurence of any other phase but the purely diluted Ga$_{1-x}$Mn$_x$N one at low-enough Mn contents 
and, second, to clearly identify the precipitates present at larger Mn incorporations. 
Indeed, the local environment is much different for a Mn either in GaMn$_3$N or Mn$_4$N clusters, both with the perovskite structure, and for a Mn 
substituting a Ga in wurtzite GaN: there are 2N as nearest-neighbours in the perovskite instead of 4N in the wurtzite, and the 
bond lengths are shorter. Moreover, the second neighbours allow one to distinguish GaMn$_3$N (8Mn+4Ga) from Mn$_4$N (12Mn). Two contrasted behaviours in EXAFS oscillations 
are observed in fig.~\ref{fig3}a, at low or high Mn contents, whereas intermediate Mn contents above 2$\%$ exhibit a gradual change from one case to the other (not shown). 
The Fourier transforms of these weighted oscillations (fig.~\ref{fig3}b) give distances to the central Mn ion that are in agreement with either the wurtzite 
or the perovskite structures, respectively. 

For a 1.5$\%$ Mn-doped GaN epilayer, the measurement can be fitted to a calculation based on a Mn random substitution of Ga in wurtzite GaN, 
including the first seven shells, which shows that the pure Ga$_{1-x}$Mn$_x$N phase is obtained. The two additional peaks evidenced at large $R$ values in fig.~\ref{fig3}b can also be reproduced if 
the first thirteen shells are taken into account in the calculation, but without considering multiple scattering paths.
However, these neglected paths contribute to the exact shape of the peaks, so that we only considered the first seven shells for the accurate fit shown in fig.~\ref{fig3}b. 
On the contrary, for a 18$\%$ Mn-doped GaN epilayer, EXAFS oscillations are strongly modified and now clearly fit 
the experimental oscillations of a GaMn$_3$N powder (note that these oscillations are perfectly in phase, up to large $k$ values). 
This demonstrates that most of the manganese ions contribute to this perovskite phase in the high-doping range.  
Moreover, it is clear from the experimental data (shift and amplitude, see fig.~\ref{fig3}a -up-) that the EXAFS oscillations of this 18$\%$ Mn-doped GaN 
epilayer do not fit the ones obtained with a Mn$_4$N powder. 
This result is also corroborated by x-ray absorption near edge structure (XANES) measurements, showing identical edges for both our highest Mn-doped GaN epilayer 
and the GaMn$_3$N powder, whereas a shift of about 2~eV is revealed for the Mn$_4$N one (not shown). 
All these features allow us to clearly identify the perovskite precipitates as due to GaMn$_3$N clusters 
and not to Mn$_4$N ones. This is particularly important since GaMn$_3$N is antiferromagnetic while Mn$_4$N is ferrimagnetic \cite{Bouchaud68,Fruchart78}.

Magnetisation measurements were performed using a MPMS-Quantum Design SQUID magnetometer, with an improved accuracy achieved using the 
reciprocating sample technique, the magnetic field being applied within the layer plane. We paid 
much attention to accurately subtract the large diamagnetic contribution of the substrate. This is particularly difficult at low temperature for the 
small paramagnetic signals measured. Therefore, we mainly 
ruled out any significant non linear contribution from the substrate and used the saturated magnetisation, as observed in high fields below 
$T\sim5$~K, to subtract the diamagnetic component. 

\begin{figure}     
\centerline{\epsfxsize= 18cm \epsfbox{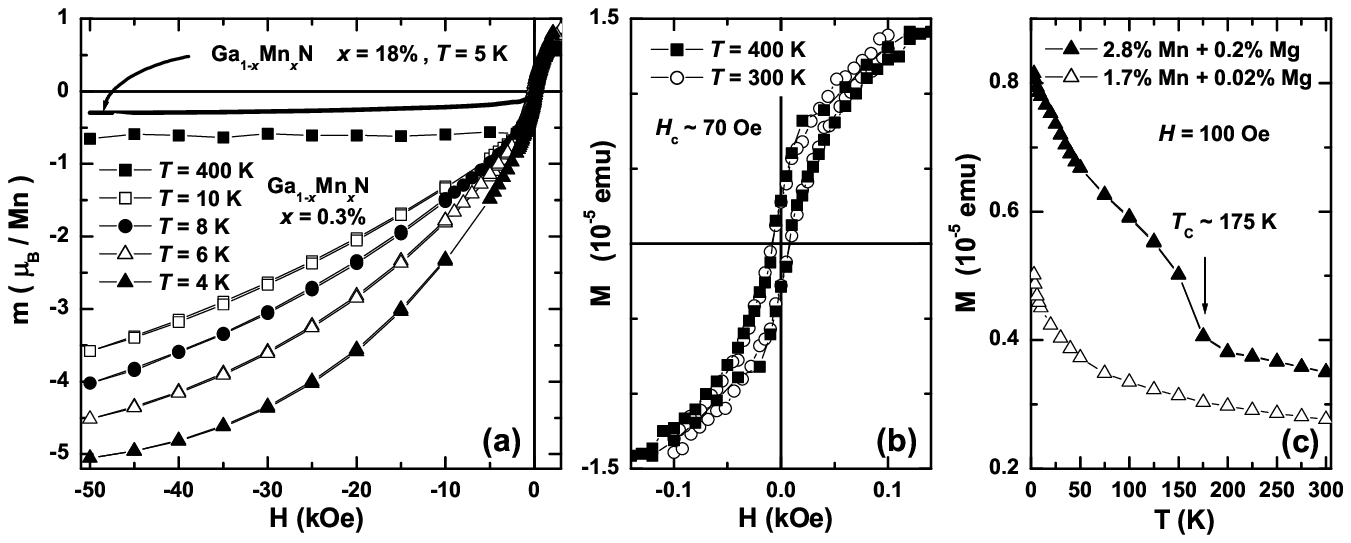}}
\caption{R. Giraud \emph{et al.}}
\label{fig4}
\end{figure}

As above, we first concentrate on the two extreme cases involving either a low Mn content (that is, below or around 2$\%$), for which only the pure diluted 
Ga$_{1-x}$Mn$_x$N phase is obtained, or the highest Mn incorporation showing a large amount of GaMn$_3$N precipitates. 
Fig.~\ref{fig4}a shows the magnetic behaviour of a Ga$_{1-x}$Mn$_x$N epilayer grown under Ga-rich conditions with $x\sim0.3\%$, that is, being far away from the smaller concentration at which additional 
precipitates were evidenced. Similarly to any other sample, a temperature-dependent paramagnetic contribution, 
being clearly observed at low temperature and in high fields, is superimposed to a nearly temperature-independent hysteretic component (see fig.~\ref{fig4}b). 
The saturated magnetisation at 4~K, of about 
5~emu/cm$^3$, gives the right order of magnitude for a Mn localized magnetic moment, that is close to 5~$\mu_B$, but does not allow to conclude quantitatively about the valence state. 
Contrary to the low Mn content case, the magnetic moment per Mn is found to be strongly reduced in a GaN:Mn epilayer containing a much larger amount of manganese ions ($\sim18\%$), 
giving less than 0.5~$\mu_B$/Mn at low temperature (solid line in fig.~\ref{fig4}a). Reduced paramagnetism shows
that most of the manganese ions ($\sim90\%$) now incorporate into GaMn$_3$N precipitates, thus corroborating EXAFS measurements, and it demonstrates 
that these clusters are large enough to have a bulk-like antiferromagnetic behaviour \cite{Bouchaud68}. 

More complicated results are expected at intermediate manganese compositions, when GaMn$_3$N precipitates are small enough to also contribute 
to the whole magnetisation, due to deviations from long-range antiferromagnetic ordering. This could lead to an optimum in the 
magnetisation \cite{Thaler02}, with some low-temperature hysteresis related to the coexistence of both Ga$_{1-x}$Mn$_x$N and very small GaMn$_3$N precipitates with a finite magnetisation. 
For instance, the early formation of such small precipitates, which cannot be resolved by x-ray analysis, is evidenced below about 10~K by a slow magnetic 
relaxation component in a $1.7\%$ Mn-doped GaN epilayer and at higher intermediate Mn contents.
Nevertheless, this contribution of the freezing of small superparamagnetic clusters remains rather small compared to any other component, 
and it is absent for $x<1.7\%$, within the magnetisation measurement accuracy (in this case less than $\sim0.1\%$ of Mn ions incorporate 
into GaMn$_3$N clusters).

The high-temperature hysteresis loop shown in fig.~\ref{fig4}b, which behaves as an in-plane ferromagnetic contribution, was not found to be specific to Mn-doped GaN epilayers. 
Indeed, a systematic investigation of the saturated magnetisation at room temperature over a broad Mn-incorporation range does not give any significant dependence with the Mn content. 
This high-temperature hysteretic artefact was carefully checked 
not to be due to the apparatus nor to the sample holder, but to the measured sample itself. Therefore, in most cases, the low-field susceptibility consists of a 1/$T$ variation with temperature, 
adding to an offset which is reminiscent of the artefact. 

However, the susceptibility of samples co-doped with Mn and Mg reveals an important new feature: a clear anomaly is observed near 175~K at large enough Mg contents, as shown in fig.~\ref{fig4}c. 
This anomaly remains very similar for Mn contents much higher than 2.8$\%$ (that is, when a large amount of GaMn$_3$N clusters is formed), as if the Mn incorporation in the diluted phase saturates (not shown). 
Note that the ferromagnetic component is observed up to $T_c\sim175$~K. Such a value of $T_c$ is large enough not to be due to unresolved 
magnetic precipitates, mainly because large blocking temperatures are ruled out in tiny clusters with a small anisotropy. 
All large manganese nitride clusters are either antiferromagnets or highly compensated ferrimagnets, which 
could explain the small amplitude of the magnetisation measured in our experiments. Nevertheless, the small change in the susceptibility anomaly at large Mn 
incorporations suggests that large clusters with the perovskite structure do not contribute significantly to the magnetisation. 
In addition, the associated N\'eel temperatures are much different than the observed values of $T_c$. The low-field magnetisation remains rather small 
at low temperature, as compared to a metallic Ga$_{1-x}$Mn$_x$As epilayer, and most of Mn impurities still remain in the paramagnetic 
state, which points to the occurence of inhomogeneous magnetic properties.  
Finally, the anomaly in the susceptibility is also independent of the artefact hysteresis loop and, as a Mg impurity usually acts as an acceptor in GaN, it suggests the possible occurence of hole-mediated ferromagnetism.

To sum up, this study shows that diluted Ga$_{1-x}$Mn$_x$N epilayers which do not contain any secondary phase, exhibit a ferromagnetic behaviour below 
room temperature, when introducing a significant amount of additional non-magnetic impurities, which usually act as acceptors in GaN. 
Further investigations of the exact role of the Mn and Mg doping levels are still required to control both the ordering temperature and to make 
ferromagnetism more homogeneous. 
In Ga$_{1-x}$Mn$_x$As, the coexistence of magnetic phases is usually attributed to a random distribution of localized Mn-impurities exchange-coupled 
to each other by a spatially-inhomogeneous carrier density, which can be expected in a system close to its Mott transition. Yet, an inhomogeneous 
distribution of Mn ions cannot be excluded. Note, however, that large band-gap ferromagnetic semiconductors are quite different from previously 
studied materials, since the $3d$-band as well as the Fermi energy are expected to lie within the gap \cite{Kulatov02,Kronik02}, and not within the 
valence band. Therefore, a better understanding of the carrier and 
Mn ion spatial distributions is necessary to highlight the origin of ferromagnetism in wide band-gap DMS. Finally, 
the conductivity of GaN:Mn layers is often found to be of $n$-type, instead of $p$-type as usually required, and previous claims for room-temperature 
ferromagnetism in such samples must be considered with caution, because of uncontrolled but reproductive hysteretic artefacts when small magnetisations of thin films are measured.

\vspace{10mm}

\centerline{Acknowledgements}
We thank Y.~Genuist for his technical assistance, P.~Hollinger, O.~Kaitasov and the PROBION company for the 
SIMS measurements, and O.~Proux and J.-L.~Hazemann for their support at the ESRF. 
J.-F. Jacquot is acknowledged for his support with SQUID measurements. 
SK thanks the Japanese Overseas Research Fellow sponsored by Monbukagakushô. 
This work was supported by the EC FENIKS project G5RD-CT-2001-00535.

\newpage

\centerline{LIST OF FIGURES CAPTIONS}

\begin{list}{FIG.~\arabic{listfig}}{\usecounter{listfig}}

\item Growth of Mn-doped GaN epilayers. (a), Ga flux dependence of the Ga desorption time, after the growth of GaN (filled squares) and 
Mn-doped GaN for two values (open triangles and circles) of the Mn-beam equivalent pressure (BEP), the N$_2$ flow and the substrate temperature being kept constant (surface stoichiometry is achieved close to 0.27~ML/s). (b), Specular RHEED intensity 
obtained during the growth of both GaN and Ga$_{1-x}$Mn$_x$N (with $x\sim1.7\%$), under N-rich conditions with a Ga flux of 0.23~ML/s. 
The indicated growth rate is deduced from the RHEED oscillations.

\item (a), X-ray $\theta$--2$\theta$ diffraction patterns of GaN:Mn epilayers. For a 18$\%$ Mn-doped GaN epilayer, the presence of oriented clusters with the perovskite structure is revealed. An additional AlN buffer layer was also grown for transport measurements. 
(b), Observation of perovskite clusters at Mn contents larger than about 2$\%$, which are shown to be GaMn$_3$N precipitates by x-ray absorption.

\item Mn local environment in GaN:Mn epilayers. (a), EXAFS oscillations as measured in the two extreme cases, that is, for a Mn content either below 2$\%$ or as large as 18$\%$. 
The latter case is compared with EXAFS oscillations obtained in GaMn$_3$N and Mn$_4$N powders, with the perovskite structure.
(b), Fourier transforms of the weighted EXAFS oscillations shown in (a), within the $k$-range [4.25, 11.7]~\AA$^{-1}$. Using the FEFITT routine package,
the calculation based on wurtzite GaN, with a central Ga atom substituted by a Mn atom, takes every interferences paths into account. Including the first seven shells, 
an accurate fit is obtained, without any additional contribution but a slight change of about 2$\%$ in the Mn-N bound length to the first 4N neighbours (see also \cite{Soo01,Sato02,Biquard03}).

\item (a), Paramagnetic behaviour in Ga$_{1-x}$Mn$_x$N with $x\sim0.3\%$. A limited magnetisation is observed at 
$T=5$~K in our 18$\%$ Mn-doped sample (see solid line). 
(b), A nearly temperature-independent hysteretic artefact is superimposed at low fields to the paramagnetic component.
(c), Temperature dependence of the field-cooled magnetisation ($H=100$~Oe) in samples co-doped with Mg impurities. 
For a large enough Mg content, an anomaly is evidenced near 175~K.

\end{list}

\newpage

\newpage

\newpage

\newpage

\end{document}